\documentclass[12pt,preprint]{aastex}

\begin{document}
\slugcomment{Submitted to ApJL: 2003--08--13}
\shorttitle{Rydberg Bands of Interstellar CO}
\shortauthors{Sheffer, Federman, \& Andersson}

\title{{\it FUSE} Measurements of Rydberg Bands of Interstellar CO
between 925 and 1150 \AA\footnotemark[1]}
\author{Y. Sheffer\altaffilmark{2}, S. R. Federman\altaffilmark{2},
and B.-G. Andersson\altaffilmark{3}}

\altaffiltext{2}{Department of Physics and Astronomy, University of Toledo, Toledo, OH 43606;
ysheffer@physics.utoledo.edu, sfederm@uoft02.utoledo.edu}
\altaffiltext{3}{Department of Physics and Astronomy, Johns Hopkins University, Baltimore, MD 21218;
bg@pha.jhu.edu}

\footnotetext[1] {Based on observations made with the NASA-CNES-CSA {\it Far Ultraviolet
Spectroscopic Explorer (FUSE)}, which is operated for NASA by the Johns Hopkins University under
NASA contract NAS5-32985.}

\begin{abstract}
We report the detection of 11 Rydberg bands of CO in {\it FUSE} spectra of the sight line toward
HD203374A. Eight of these electronic bands are seen in the interstellar medium for the first time.
Our simultaneous fit of five non-Rydberg {\it A$-$X} bands
together with the strongest Rydberg band of CO, {\it C$-$X} (0$-$0), yields a 4-component
cloud structure toward the stellar target. With this model we synthesize the other
Rydberg bands in order to derive their oscillator strengths. We find that the strength of some
bands was underestimated in previously published results from laboratory measurements. The
implication is important for theoretical calculations of the abundance of interstellar CO, because
its dissociation and self shielding depend on oscillator strengths for these bands.
\end{abstract}


\section{Introduction}

The second most abundant molecule in interstellar clouds, CO, has a rich and long observational
history in the radio, near-infrared, and far-ultraviolet (FUV) regimes. This sequence of spectral
windows provides a view of pure-rotational and rotational-vibrational transitions among
ground-state ($X~^1\Sigma^+$) levels, and electronic transitions between the $X$ state and the
first excited singlet state, $A~^1\Pi$ \citep{mn94}. Deeper into the FUV, additional prominent
dipole-allowed bands of CO appear, involving the Rydberg states $B~^1\Sigma^+$, $C~^1\Sigma^+$, and
$E~^1\Pi$. First astronomical observations of the {\it C$-$X} (0$-$0) band at 1088 \AA, which has
the largest band oscillator strength ($f_{00}$ = 0.123) according to \citet[hereafter F01]{f01}, and of the
{\it E$-$X} (0$-$0) band at 1076 \AA\ were conducted with the {\it Copernicus} satellite with a
spectral resolution ($R$) of 20,000 \citep{jenk}. Later, \citet{mort} also reported {\it Copernicus}
data on the {\it B$-$X} (0$-$0) band at 1150 \AA. This band is
only dipole-allowed band other than the {\it A$-$X} bands that has been fully resolved into
individual rotational lines -- see the GHRS spectrum from the {\it Hubble Space Telescope}
({\it HST}) in Figure 5 of \citet{sh92}. Another UV space
mission following {\it Copernicus} that detected the {\it C$-$X} (0$-$0) band was the
{\it Orbiting and Retrievable Far and Extreme Ultraviolet Spectrometer} ({\it ORFEUS}), albeit
at the more modest resolution of $R$ = 3,000 \citep{lee}.

In 1999, {\it FUSE} was launched, covering
the spectral range of 905$-$1187 \AA\ at the same resolution as that of {\it Copernicus}
\citep{moos,sahn}. In addition to covering the {\it B$-$X}, {\it C$-$X}, and {\it E$-$X} bands,
{\it FUSE} should be able to detect a significant number of bluer Rydberg bands below 1075 \AA.
These are the signatures of high-lying states of CO that converge onto
electronic states in CO$^+$. The Rydberg states have an important role in the photodissociation
of CO, which occurs via line absorption of FUV photons from the ground state, followed by
predissociation into atomic C and O. \citet{letz}
mapped 46 bands between 885$-$1099 \AA, i.e., to the blue of the {\it B$-$X} (0$-$0) and (1$-$0)
bands. Of these, 43 bands are predissociating, a fact that enabled \citet{vial} to revise the rate
of CO photodissociation upward by a factor of 20. Not only was CO found to be more readily
destroyed when exposed to the UV field, but the nature of line absorption leads to strong self
shielding of this molecule as its column density increases into a molecular cloud
\citep{vdb,wari}. Self shielding limits the effectiveness of photodissociation. Therefore, any
revisions to $f$-values for the Rydberg bands are an important input for theoretical models of
the interstellar CO abundance.

These bands have been studied in the laboratory. Line lists and band constants were presented
by \citet{e90}, \citet[hereafter E91]{e91}, and \citet{e92}. Unfortunately, it seems that optical
depth
effects hindered reliable measurements of $f$-values for many of these bands, which are invariably
strong. Later investigations at high resolution and low temperature (20 K) were reported by Stark,
Yoshino, and collaborators \citep[and references therein]{yosh}. In a recent study F01
measured laboratory $f$-values for five bands involving the B, C, and E states and found that the
$f$-values were larger than those
in E91 by factors between 1.08 (a negligible difference) and 1.99 (a significant
one). Clearly, there is a need to extend the re-evaluations to other Rydberg bands. {\it FUSE}
spectra turn out to be complementary to the laboratory measurements, since they allow internally
self-consistent $f$-values to be extracted from interstellar CO bands. Here we present the
interstellar results for CO toward \objectname{}{HD203374A}, including eight Rydberg bands
that are detected in the ISM for the first time.

\section{Observations and Analysis}

{\it FUSE} observations of HD203374A were obtained on 2001 August 2, as part of the {\it FUSE}
Guest Investigator program B030. The target, which is a member of the Cep OB2 Association, is
a B0 IV star with a reddening of $E_{B-V}$ = 0.6 mag. Thirty four sub-exposures were secured in
the HIST (spectral image) mode, totaling 18,653 s. All sub-exposures were co-added and calibrated
using the CalFUSE software (version 2.4). The data were then re-binned by a factor of four to yield
two pixels per nominal resolution element (0.05 \AA). The binned pixels are of higher
signal-to-noise (S/N) thanks to the large number of sub-exposures, which effectively diluted the
fixed pattern noise of {\it FUSE}. Scattered light was successfully corrected in the CalFUSE
processing, with the opaque flat cores of H$_2$ lines found to be well below the 1\% normalized
flux level. Most of the {\it FUSE} spectral range is covered by two (sometimes three) independent
detector segments, except for the 1017$-$1075 \AA\ region, where four segments are available.
Each of the CO bands was modeled in each individual segment, as well as in the final co-added
spectrum from all available segments. The spectral resolution varies between detectors:
$R$ = 15,000 for bands in the SiC channels and 19,000 in the LiF ones. The only band
with equal contributions from both types, {\it E$-$X} (1$-$0), was modeled with $R$ = 17,000.
The final co-added spectra that become
input for our synthesis code, {\sc ismod}, have a S/N ratio of 80 to 100 per pixel. {\sc ismod}
fits each rotational line with a Voigt profile whose Gaussian component is based on the Doppler
broadening parameter, and whose Lorentzian part is controlled by the inverse lifetime as
determined by radiative decay and predissociation. Upper level lifetimes, which were treated as
fixed input, were gathered from E91, \citet{mn94}, \citet{ubac}, and \citet{cacc}.
The code then minimizes the root-mean-square residuals between the data and a model that
incorporates freely-varying parameters, using shrinking step sizes down to 10$^{-4}$ of
a parameter's value. Although {\sc ismod} was first used to synthesize CO bands in {\it HST}
spectra, we recently modified it to synthesize H$_2$ bands seen by {\it FUSE}. This new derivative
of {\sc ismod} was essential for the initial step of dividing out the H$_2$ lines before the
blended CO bands could be recovered and then modeled successfully.

\section{Results}

As a prelude to CO analysis, a model of the H$_2$ gas toward HD203374A was derived. It shows
a total H$_2$ column density of 5.10 ($\pm$ 0.05) $\times$ 10$^{20}$ cm$^{-2}$ and a kinetic
temperature (or $T_{1,0}$ of H$_2$) of 74.5 $\pm$ 1.5 K. This model was varied by small steps to
account for {\it FUSE} spectral changes in various wavelength regions and was then divided into CO
spectral cuts as a means of deblending the Rydberg bands. Each CO band involves R, P, and for upper
$\Pi$ states Q transitions from four significantly populated rotational levels ($J^\prime$$^\prime$
$\leq$ 3) in the ground state. Bands that involve an upper $\Sigma^+$ state, e.g., {\it B$-$X}
(1$-$0) and {\it K$-$X} (0$-$0), resemble a blended double absorption profile from their P and R
branches (see Figure 1). Those that involve a $\Pi$ state, such as {\it E$-$X} (1$-$0) and
{\it W$-$X} (2$-$0), have a Q branch containing a pileup of lines between the R and P branches,
and thus appear as a single feature. While {\it FUSE} spectra cannot resolve individual rotational
lines, a splitting of each transition into four cloud components, as described below, is required
to synthesize the bands in an adequate manner.

The derivation of $f$-values for optically-thick CO bands requires that both the total CO column
density, $N$(CO), and the cloud structure are known. Five high-resolution ($R$ = 140,000 after
$\times$2 binning) spectral cuts from {\it HST}/STIS data toward HD203374A were kindly supplied
to us by E. Jenkins. These include five consecutive {\it A$-$X} bands of CO, from (7$-$0) at
1344 \AA\ to (11$-$0) at 1263 \AA, for which accurately known $f$-values are available
\citep{chan,lamb}. Although the weaker bands are adequate for $N$(CO) derivations, we also
incorporated the strong {\it C$-$X} (0$-$0) band from our {\it FUSE} data as a calibrator of this
analysis. Such a strong band is especially sensitive to variables that control the optical depth,
i.e., the position, fraction, and width of individual clouds, as well as the excitation temperature
($T_{ex}$) of ground rotational levels.
Together with the {\it C$-$X} (0$-$0) band, the run of
bands covers a range of 2.8 orders of magnitude in $f$-value, thereby being able to accommodate
any other FUV band as part of a robust model fit. Note that {\it C$-$X} (0$-$0) is a bit
problematic, owing to a blending of its P branch with the strong interstellar line of \ion{Cl}{1}
at 1088.06 \AA\ \citep{f86}. However, the important benefit of using this band was realized by
limiting model fitting to its R branch.

Our synthesis was based on the four-component cloud structure seen in high-resolution
($R$ = 200,000) CH spectra at visible wavelengths \citep{pan}. The structure was kept fixed during
H$_2$ synthesis, but was allowed to vary for the high-resolution CO fits. The resulting relative
velocity (in km s$^{-1}$), fractional strength, and Doppler width
($b$-value = $\Delta v$(FWHM)/1.665, in km s$^{-1}$) are (0.0, 0.44, 0.77), ($-$3.6, 0.51, 0.54),
($-$7.3, 0.02, 0.59), and ($-$11.4, 0.03, 0.54). Jointly, these four components give a total
$N$(CO) of 2.41 ($\pm$ 0.24) $\times$ 10$^{15}$ cm$^{-2}$ in front of HD203374A. The fitted
equivalent widths ($W_\lambda$) for the {\it A$-$X} (7$-$0) to (11$-$0) bands are 50.9 $\pm$ 0.8,
36.3 $\pm$ 0.5, 22.6 $\pm$ 0.6, 11.7 $\pm$ 0.5, and 5.6 $\pm$ 0.7 m\AA, respectively. The
$W_\lambda$ for the {\it C$-$X} (0$-$0) band is 139 $\pm$ 13 m\AA. Modeled values
of $T_{ex}$ are 3.2, 3.5, and 5.0 K for the first three rotational levels, relative to
$J^\prime$$^\prime$ = 0. As is the case along previously studied lines of sight, CO-bearing cloud
components have the narrowest line widths among either atomic or molecular species, as well as
some of the lowest $T_{ex}$ values. At first glance it would seem that the last two components,
which have fractional strengths of $\leq$ 3\%, are too weak to be of significance in the fit.
However, for bands with large $f$-values, the
strongest components are very optically thick, allowing the weaker components to determine
band profile shapes.  

Following the simultaneous fit of the six bands, we re-fitted each of them separately in order
to gauge the associated uncertainties. Using the newly derived cloud structure and $T_{ex}$,
individual $N$(CO) values were up to 11\% away from their average. Therefore, $N$(CO) toward
HD203347A is uncertain by $\sim$10\%. Additional testing is available through comparison of three
other bands studied by F01. {\it B$-$X} (0$-$0) has fitted $W_\lambda$ = 50 $\pm$ 3 m\AA\
and $f$-value = 5.7 ($\pm$ 0.9) $\times$ 10$^{-3}$, and {\it E$-$X} (0$-$0) has fitted
values of 97 $\pm$ 2 m\AA\ and 63 ($\pm$ 7) $\times$ 10$^{-3}$, respectively. The
values for {\it B$-$X} (1$-$0), which is a first ISM detection, are presented in Table 1. 
The corresponding deviations from the $f$-values of F01 are $-$15\%, $-$8\%,
and +33\%, respectively, thus averaging
about $\pm$ 2$\sigma$ relative to our uncertainty for $N$(CO). With uncertainties of 10\% for
the {\it A$-$X} bands (Chan et al. 1993) and 13\% for {\it C$-$X} (0$-$0) (F01), the absolute
1$\sigma$ uncertainties on our $f$-values are close to 15\%.

Seven additional Rydberg bands farther into the FUV, which were not studied by F01, are
seen in our {\it FUSE} spectra (see Figure 1 and Table 1). Of these, {\it K$-$X} (0$-$0),
{\it W$-$X} (1$-$0), and {\it W$-$X} (2$-$0) are least affected by blending with H$_2$ lines.
Indeed, these three bands have appreciable contributions to the photodissociation rate of
interstellar CO, as can be seen in Table 2 of \citet{vdb}. Since $N$(CO) is now known,
self-consistent $f$-values for all {\it FUSE} bands can be derived. Ratios of interstellar
$f$-value to that of E91 are 1.27 $\pm$ 0.24 and 1.75 $\pm$ 0.26 for {\it B$-$X} (0$-$0)
and {\it E$-$X} (0$-$0), respectively. Together with the $f$-value ratios given in Table 1,
seven bands out of 10 are stronger by factors between 1.07 and 1.75
relative to E91, although most of these factors are smaller than their 1$\sigma$
uncertainties. The three bands with the most significant factors are {\it E$-$X} (0$-$0),
{\it K$-$X} (0$-$0), and {\it B$-$X} (1$-$0), which have $f$-value ratios of 1.75 $\pm$ 0.26
(deviating by 2.9$\sigma$), 1.48 $\pm$ 0.24 (2.0$\sigma$), and 1.47 $\pm$ 0.21 (2.2$\sigma$),
respectively. For the weakest of these bands, {\it E$-$X} (1$-$0), most determinations are
consistent with an $f$-value of $\sim$3.5 $\times$ 10$^{-3}$.

The new interstellar band $f$-values confirm the conclusions of F01, namely, that the
older $f$-values in E91 are too small for moderately-strong to strong bands. However, the
average correction for the bands to the blue of {\it E$-$X} (0$-$0) is not as large as the one
derived from the redder bands of F01, which was dominated by the strongest bands of CO,
{\it C$-$X} (0$-$0) and {\it E$-$X} (0$-$0). Therefore, the trend of larger corrections for
stronger bands from F01 is also confirmed. Rotational lines of Rydberg bands below
$\sim$1000 \AA\ tend to have larger natural widths owing to level lifetimes that are
predominantly shorter than $\sim$10$^{-11}$ s, whereas above $\sim$1000 \AA\, longer lifetimes are
encountered. The larger natural widths result in smaller optical depths at line centers, thus
alleviating the line saturation more readily associated with redder bands of similar oscillator
strength. This may explain why interstellar $f$-values agree better with synchrotron absorption
results for bands below 970 \AA.

There is one exception to this trend. The band {\it L$^\prime-$X} (1$-$0) has a ratio
of 0.81 $\pm$ 0.12 (1.6$\sigma$) relative to Eidelsberg et al.'s (1991) result. As shown in
Table 1, the 20 K result from \citet{s93} is somewhat smaller than ours. Two earlier studies found
that {\it L$^\prime-$X} (1$-$0) is weaker than {\it L$-$X} (0$-$0), a band only 0.6 \AA\ to the
blue of {\it L$^\prime-$X} (1$-$0), and blended with it at room temperature \citep{letz,s91}.
The roles are reversed in E91, with {\it L$^\prime-$X} (1$-$0) being the stronger band,
thus prompting us to suspect some contamination from {\it L$-$X} (0$-$0). Note that the continuum
under this interstellar band had to be corrected by 10\% owing to unidentified (probably stellar)
absorption. Without this correction, the $f$-value would be 1.16 $\pm$ 0.16 relative to E91.

The band {\it I$^\prime-$X} (0$-$0) is a special case. Initial inspection showed that only a weak
feature in the {\it FUSE} spectra coincided with the E91 position of the band, while
a much larger feature to the blue could not be identified. In the laboratory, this
band is very diffuse, and shows at least three intensity peaks in its room-temperature profile,
\citep[their Figure 1]{s91}. Such a structure is in conflict with the E91 tabulation that
provides two branches (R and P) and the assignment $\Sigma^+-\Sigma^+$ for this band. We suspected
that {\it I$^\prime-$X} (0$-$0) suffers from strong perturbations, or that it may be blended with
another, hitherto unaccounted for, CO band. A laboratory spectrum acquired at 80 K (M. Eidelsberg
2003, private communication) more clearly defines the R and P branches of {\it I$^\prime-$X}
(0$-$0), leading to a shift of $-$0.243 \AA\ in line positions relative to E91. Since this
shift leads to excellent agreement with the larger feature in the {\it FUSE} spectra, we adopted
this shift in our synthesis of the band, yielding the $f$-value given in Table 1. The agreement
with laboratory results is excellent. This is the only case were we did allow the natural width
of the lines to vary, owing to the extremely diffuse nature of the band. Our fit of its wings
resulted in a lifetime change from 10$^{-12}$ s (E91) to 4.2 ($\pm$ 0.8) $\times$ 10$^{-13}$
s. For a complementary confirmation it would be highly desirable to obtain a new laboratory
recording of this region at a very low $T_{ex}$.

\section{Concluding Remarks}

In this paper we showed that {\it FUSE} spectra can be utilized to measure the Rydberg bands of CO
by isolating them from the strong presence of H$_2$. Indeed, the bands not totally obscured are
precisely those that play the major role in CO photodissociation. Thus observations via
{\it FUSE} have a direct impact on detailed modeling of the CO distribution in the ISM. {\it FUSE}
can be turned into a reliable estimator of CO abundance and conditions, provided the number of
velocity components is available from high-resolution ground-based spectra
of CH, or from {\it HST}/STIS data of CO itself, i.e., its {\it A$-$X} bands. 

Most of the interstellar results confirm the larger $f$-values of relatively strong Rydberg bands of
CO reported in recent laboratory studies (see F01). Interestingly, this is especially true for the
{\it B$-$X} (0$-$0), {\it E$-$X} (0$-$0), and the first three bands in Table 1, but much less
obvious for all bands below 970 \AA, the result of diminishing lifetimes for bluer bands. Overall,
interstellar values show good agreement with laboratory $f$-values obtained at 20 K
\citep{s93,yosh} and with laser absorption spectroscopy at very high resolution \citep{s99}.
Otherwise, room-temperature spectroscopy suffers from serious blending of bands,
especially below $\sim$1000 \AA.

In an earlier paper \citep{sh02} we showed that the older theoretical models of \citet{vdb} and
Warin et al. (1996) reproduced either the observed isotopomeric ratios or the observed column
densities of four CO isotopomers, but a given model could not reproduce both quantities. The new
results from CO toward HD203374A and the recent laboratory determinations of band $f$-values
provide evidence that CO is more resilient to destruction by enhanced self shielding because larger
$f$-values cause higher optical depths for individual lines. Theoretical models of molecular
clouds that explore kinematic aspects of the CO photodissociation rate, e.g., \citet{roll}, will
also be affected. The new $f$-values of Rydberg bands in the FUV should be incorporated into the
next generation of models to yield improved predictions of CO abundance.

\acknowledgments

We thank NASA for grants NAG5-8961, NAG5-10305, and NAG5-11440. We are grateful to Dr. E. Jenkins
for sharing STIS data, to Dr. M. Eidelsberg for an update of the laboratory view of {\it I$^\prime$}(0),
and to Dr. K. Pan for providing us with the cloud structure from CH spectra.

\begin{deluxetable}{lcccccccc}
\rotate
\tabletypesize{\scriptsize}
\tablewidth{0pt}
\tablecaption{Comparison of Rydberg Band $f$-value Measurements\tablenotemark{a}}
\tablehead{
\colhead{Parameter/Band}
&\colhead{{\it B}1}
&\colhead{{\it E}1}
&\colhead{{\it K}0}
&\colhead{{\it L$^\prime$}1}
&\colhead{{\it W}1}
&\colhead{{\it W}2}
&\colhead{{\it I$^\prime$}0}
&\colhead{{\it W}3}
}
\startdata
$\lambda_0[R(0)]$ (\AA) &1123.57 &1051.67 &970.33 &968.85 &956.22 &941.15 &940.17 &925.79 \\
{\it FUSE} $W_\lambda$ (m\AA) &19 $\pm$ 2 &37 $\pm$ 6 &104 $\pm$ 4 &95 $\pm$ 4 &117 $\pm$ 5 &109 $\pm$ 7 &259 $\pm$ 13 &125 $\pm$ 5 \\
\underline{$f$-values (10$^{-3}$)\tablenotemark{b,c}} \\
E91 &0.72 $\pm$ 0.07 &2.5 $\pm$ 0.3 &21.0 $\pm$ 2.1 &12.4 $\pm$ 1.2\tablenotemark{d} &13.5 $\pm$ 1.4 &25.8 $\pm$ 2.6 &21.3 $\pm$ 2.1 &16.3 $\pm$ 1.6 \\
S91,S92,S93 &\nodata &3.0 $\pm$ 0.3 &{\bf 26.8 $\pm$ 3.8} &{\bf 7.5 $\pm$ 0.8} &14.8 $\pm$ 1.5 &30.0 $\pm$ 3.0 &23.6 $\pm$ 2.4 &14.9 $\pm$ 1.5 \\
C93 &1.32 $\pm$ 0.13 &3.53 $\pm$ 0.35 &\nodata &\nodata &\nodata &\nodata &\nodata &\nodata \\
Y95 &\nodata &\nodata &{\bf 33.5 $\pm$ 5.0} &\nodata &\nodata &{\bf 20.4 $\pm$ 3.1} &\nodata &{\bf 17.0 $\pm$ 2.6} \\
Z97 &\nodata &4.67 $\pm$ 0.66 &\nodata &\nodata &\nodata &\nodata &\nodata &\nodata \\
S99 &1.1 $\pm$ 0.1 &\nodata &\nodata &\nodata &\nodata &\nodata &\nodata &\nodata \\
F01 &0.80 $\pm$ 0.12 &\nodata &\nodata &\nodata &\nodata &\nodata &\nodata &\nodata \\
HD203374A &1.06 $\pm$ 0.11 &3.3 $\pm$ 1.1 &31 $\pm$ 4 &10.1 $\pm$ 1.1 &15.8 $\pm$ 2.0 &23 $\pm$ 5 &22.6 $\pm$ 2.9 &19.8 $\pm$ 2.4 \\
\\
ISM/Lab\tablenotemark{e} &1.47 $\pm$ 0.21 &1.32 $\pm$ 0.46 &1.48 $\pm$ 0.24 &0.81 $\pm$ 0.12 &1.07 $\pm$ 0.17 &0.88 $\pm$ 0.21 &1.06 $\pm$ 0.17 &1.21 $\pm$ 0.19 \\
\enddata
\tablenotetext{a}{Band notation has been shortened by deleting the ground state.}
\tablenotetext{b}{Bold face results are laboratory measurements at 20 K.}
\tablenotetext{c}{References are: (E91) Eidelsberg et al. 1991; (S91,S92,S93) Stark et al.
1991, 1992, 1993; (C93) Chan et al. 1993; (Y95) Yoshino et al. 1995; (Z97) Zhong et al. 1997;
(S99) Stark et al. 1999; (F01) Federman et al. 2001; (HD203374A) this paper.}
\tablenotetext{d}{Contamination from {\it L}0 suspected; see text.}
\tablenotetext{e}{The $f$-value ratio of this paper to \citet{e91}.}
\end{deluxetable}

\begin{figure}
\epsscale{0.4}
\plotone{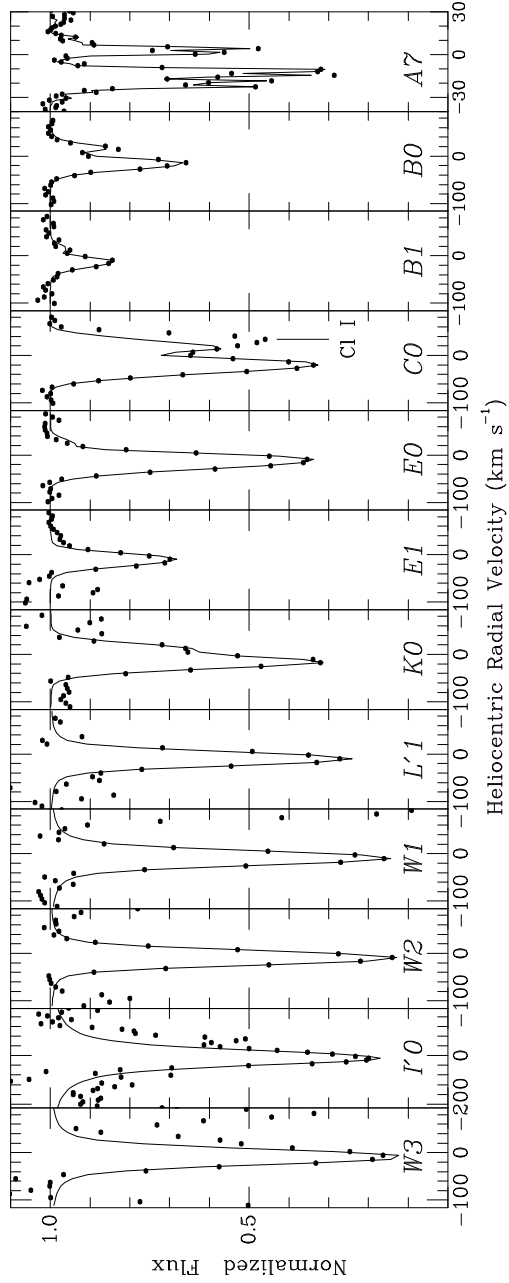}
\caption{Montage of 11 Rydberg bands of CO from {\it FUSE}, together with a sample Fourth Positive
band, {\it A$-$X} (7$-$0), from {\it HST}, toward HD203374A. Band notation has been shortened by
deleting the ground state. The sequence of bands follows their wavelengths order, with redder bands
to the right. The heliocentric radial velocity for the $R$(0) line of all bands has been
adopted from the STIS spectra, owing to the uncalibrated {\it FUSE} radial velocity scale. The
scale is expanded for $A$7 and is reduced for the broad $I^{\prime}$0 band.}
\end{figure}

\end{document}